\input harvmac

\lref\StromingerPC{
A.~Strominger,
``Open string creation by S-branes,''
arXiv:hep-th/0209090.
}

\lref\GutperleXF{
M.~Gutperle and A.~Strominger,
``Timelike boundary Liouville theory,''
Phys.\ Rev.\ D {\bf 67}, 126002 (2003)
[arXiv:hep-th/0301038].
}

\lref\OkudaYD{
T.~Okuda and S.~Sugimoto,
``Coupling of rolling tachyon to closed strings,''
Nucl.\ Phys.\ B {\bf 647}, 101 (2002)
[arXiv:hep-th/0208196].
}

\lref\ConstableRC{
N.~R.~Constable and F.~Larsen,
``The rolling tachyon as a matrix model,''
JHEP {\bf 0306}, 017 (2003)
[arXiv:hep-th/0305177].
}

\lref\OkuyamaJK{
K.~Okuyama,
``Comments on half S-branes,''
JHEP {\bf 0309}, 053 (2003)
[arXiv:hep-th/0308172].
}

\lref\KlebanovKM{
I.~R.~Klebanov, J.~Maldacena and N.~Seiberg,
``D-brane decay in two-dimensional string theory,''
JHEP {\bf 0307}, 045 (2003)
[arXiv:hep-th/0305159].
}

\lref\SenSM{
A.~Sen,
``Tachyon condensation on the brane antibrane system,''
JHEP {\bf 9808}, 012 (1998)
[arXiv:hep-th/9805170].
}

\lref\LambertZR{ N.~Lambert, H.~Liu and J.~Maldacena, ``Closed
strings from decaying D-branes,'' arXiv:hep-th/0303139.
}

\lref\MukhopadhyayEN{ P.~Mukhopadhyay and A.~Sen, ``Decay of
unstable D-branes with electric field,'' JHEP {\bf 0211}, 047
(2002) [arXiv:hep-th/0208142].
}

\lref\ReyXS{
S.~J.~Rey and S.~Sugimoto,
``Rolling tachyon with electric and magnetic fields: T-duality approach,''
Phys.\ Rev.\ D {\bf 67}, 086008 (2003)
[arXiv:hep-th/0301049].
}

\lref\ChenFP{
B.~Chen, M.~Li and F.~L.~Lin,
``Gravitational radiation of rolling tachyon,''
JHEP {\bf 0211}, 050 (2002)
[arXiv:hep-th/0209222].
}

\lref\SenTM{
A.~Sen,
``Dirac-Born-Infeld action on the tachyon kink and vortex,''
Phys.\ Rev.\ D {\bf 68}, 066008 (2003)
[arXiv:hep-th/0303057].
}

\lref\McGreevyKB{
J.~McGreevy and H.~Verlinde,
``Strings from tachyons: The c = 1 matrix reloaded,''
JHEP {\bf 0312}, 054 (2003)
[arXiv:hep-th/0304224].
}

\lref\DouglasUP{
M.~R.~Douglas, I.~R.~Klebanov, D.~Kutasov, J.~Maldacena,
E.~Martinec and N.~Seiberg,
``A new hat for the c = 1 matrix model,''
arXiv:hep-th/0307195.
}

\lref\TakayanagiSM{
T.~Takayanagi and N.~Toumbas,
``A matrix model dual of type 0B string theory in two dimensions,''
JHEP {\bf 0307}, 064 (2003)
[arXiv:hep-th/0307083].
}

\lref\YeeEC{
H.~U.~Yee and P.~Yi,
``Open / closed duality, unstable D-branes, and coarse-grained closed
strings,''
Nucl.\ Phys.\ B {\bf 686}, 31 (2004)
[arXiv:hep-th/0402027].
}
\lref\YiHD{
P.~Yi,
``Membranes from five-branes and fundamental strings from Dp branes,''
Nucl.\ Phys.\ B {\bf 550}, 214 (1999)
[arXiv:hep-th/9901159].
}

\lref\BergmanXF{
O.~Bergman, K.~Hori and P.~Yi,
``Confinement on the brane,''
Nucl.\ Phys.\ B {\bf 580}, 289 (2000)
[arXiv:hep-th/0002223].
}

\lref\NagamiYZ{ K.~Nagami, ``Closed string emission from unstable
D-brane with background electric
field,''
JHEP {\bf 0401}, 005 (2004) [arXiv:hep-th/0309017].
}

\lref\KwonQN{
O.~K.~Kwon and P.~Yi,
``String fluid, tachyon matter, and domain walls,''
JHEP {\bf 0309}, 003 (2003)
[arXiv:hep-th/0305229].
}

\lref\SenNU{
A.~Sen,
``Rolling tachyon,''
JHEP {\bf 0204}, 048 (2002)
[arXiv:hep-th/0203211].
}

\lref\SenIN{
A.~Sen,
``Tachyon matter,''
JHEP {\bf 0207}, 065 (2002)
[arXiv:hep-th/0203265].
}

\lref\SenAN{
A.~Sen,
``Field theory of tachyon matter,''
Mod.\ Phys.\ Lett.\ A {\bf 17}, 1797 (2002)
[arXiv:hep-th/0204143].
}

\lref\SenKD{
A.~Sen,
``Fundamental strings in open string theory at the tachyonic vacuum,''
J.\ Math.\ Phys.\  {\bf 42}, 2844 (2001)
[arXiv:hep-th/0010240].
}

\lref\SenXS{
A.~Sen,
``Open-closed duality at tree level,''
Phys.\ Rev.\ Lett.\  {\bf 91}, 181601 (2003)
[arXiv:hep-th/0306137].
}

\lref\GutperleAI{
M.~Gutperle and A.~Strominger,
``Spacelike branes,''
JHEP {\bf 0204}, 018 (2002)
[arXiv:hep-th/0202210].
}

\lref\GibbonsHF{
G.~W.~Gibbons, K.~Hori and P.~Yi,
``String fluid from unstable D-branes,''
Nucl.\ Phys.\ B {\bf 596}, 136 (2001)
[arXiv:hep-th/0009061].
}

\lref\GibbonsTV{
G.~Gibbons, K.~Hashimoto and P.~Yi,
``Tachyon condensates, Carrollian contraction of Lorentz group, and fundamental
strings,''
JHEP {\bf 0209}, 061 (2002)
[arXiv:hep-th/0209034].
}

\lref\HwangAQ{
S.~Hwang,
``No Ghost Theorem For SU(1,1) String Theories,''
Nucl.\ Phys.\ B {\bf 354}, 100 (1991).
}

\lref\EvansQU{
J.~M.~Evans, M.~R.~Gaberdiel and M.~J.~Perry,
``The no-ghost theorem for AdS(3) and the stringy exclusion principle,''
Nucl.\ Phys.\ B {\bf 535}, 152 (1998)
[arXiv:hep-th/9806024].
}

\Title {\vbox{ \baselineskip12pt
\hbox{hep-th/0409050}\hbox{UCLA/04/TEP-36}\hbox{KIAS-P04036}
}}
{\vbox{
\centerline{
Winding Strings and Decay of  D-Branes with Flux
}}}

\centerline{ Michael Gutperle\foot{gutperle@physics.ucla.edu}}
\smallskip
\centerline{ Department of Physics and Astronomy, UCLA, Los Angeles,
CA 90095-1547, USA}
\bigskip
\centerline{and}
\bigskip
\centerline{Piljin Yi\foot{piljin@kias.re.kr}}
\smallskip
\centerline{  Institute for Advanced
Study, Princeton, NJ 08540, USA}\centerline{and}
\centerline{ School of Physics, Korea Institute for Advanced
Study}
\centerline{207-43, Cheongryangri-Dong, Dongdaemun-Gu,
Seoul 130-722, Korea}

\vskip .3in \centerline{\bf Abstract}

\smallskip\noindent
We study the boundary state associated with the decay of an unstable
D-brane with  uniform electric field, $1 > e >0$ in the string units.
Compactifying the D-brane along the direction of the electric
field, we find that the decay process is dominated by production
of closed strings with some winding numbers; closed strings produced
are such that the winding mode carries precisely the fraction $e$
of the individual string energy. This supports the conjecture that the
final state at tree level is composed of winding strings with
heavy oscillations turned on. As a corollary, we argue that the
closed strings disperse into spacetime at a much slower rate than
the case without electric field.

\vfill

\smallskip

\newsec{Closed Strings and Unstable D-Branes with Electric Field  }

Decay of unstable D-branes \SenSM\hskip 1mm has been studied
extensively in the classical limit of open string theory yet our
understanding remains limited.
Most notably, whether and precisely how closed string degrees
of freedom emerge from the decay remains an issue. For some of
effort in searching for fundamental strings in the low
energy pictures can be found in \YiHD\BergmanXF\GibbonsHF\SenKD.
Only in the limited setting of $1+1$ dimensions, a successful recovery
of closed strings has been seen, thanks to reinterpretation of
old $c=1$ matrix model as theory of many unstable D-branes
\McGreevyKB\TakayanagiSM\DouglasUP\KlebanovKM.

For more generic string theories,
an important breakthrough came about two years ago, when A. Sen
wrote down time-dependent boundary states describing  decay of
unstable D-branes \SenNU\SenIN. This set a stage for
more precise questions about evolution of unstable D-branes
in fully stringy setting \GutperleAI\MukhopadhyayEN
\ReyXS\ChenFP\GutperleXF\OkudaYD\ConstableRC\LambertZR .
So far, however, many
 successful efforts in this regard
have been aimed at
understanding the homogeneous final state. For questions
related to dynamical objects, such as lower dimensional D-branes,
the low energy approach still proves much more powerful and
also somewhat unexpectedly accurate.\foot{Perhaps one of
the more beautiful examples of this can be found in the construction
of low energy dynamics of solitonic D-branes, {\it starting with}
the severely truncated low energy approximation of unstable
D-branes \SenTM.}

One unusual property of this tachyonic system is that  all
fundamental degrees of freedom one starts with, namely those
associated with open strings, will have to disappear eventually.
This seems to raise a paradox since the dynamics involves no
dissipation: all conserved quantities such as energy, momentum,
and gauge charges must survive somehow. In particular, it is
important to note that this has to be true even in the classical
limit, $g_s\rightarrow 0$. We are accustomed
to the idea that open string theory is self-contained at classical
level and does not require presence of closed strings, however,
and we are thus faced with quandary of where to attribute these conserved
quantities after disappearance of all dynamical degrees of freedom
available at tree level.

Through study of low energy approximation \GibbonsHF\SenAN\GibbonsTV,
and also study of boundary states \SenNU\SenIN,
both at tree level, some effective
degrees of freedom have been found to carry these conserved
quantities. They are called  tachyon matter \SenIN\ and
string fluid \GibbonsHF. Both carry energy and momentum, while the
latter carry in addition the fundamental string charge induced
by the electric
field on the worldvolume. They can both be viewed as  fluid
in that their state is specified completely by the distribution
and the flow
of energy, momentum, and  electric flux. Their equation of states
are such that no pressure is present other than the tension
along the flux lines \GibbonsHF\GibbonsTV\KwonQN.

For instance, the Hamiltonian for
the tachyon matter (in the absence of string fluid) collapses
to \SenAN
\eqn\tachyon{{\cal H}=\sqrt{(\pi_T^2)(1+(\vec\partial T)^2)},}
where  $\pi_T$ is the conjugate momentum to $T$. The resulting dynamics
is that of a perfect fluid with  density distribution
$\pi_T$ and the rotationless velocity field
$\partial_i T$, $i=1,\dots,p,$ on the unstable D$p$-brane worldvolume.
While tachyon matter is intriguing on its own, life becomes
more interesting when electric field is turned on. The final
state of the decay encodes  fundamental
string charge in the conjugate momentum $\pi_i$ of the gauge
field, which is nothing but the conserved electric flux obeying
the Gauss law, $\partial_i \pi_i=0$.  The
Hamiltonian with electric field is also pretty simple \GibbonsHF,
\eqn\string{{\cal H}=\sqrt{(\vec\pi)^2 +\pi_T^2 +(\vec P\,)^2 +
(\pi_i\partial_i T)^2},}
with  the Poynting vector $P_i=F_{ik}\pi_k+\partial_iT\pi_T$.
Again the equation of motion from this is fluidic, where the
tachyon matter density, $\pi_T$, interacts with the string fluid
density, $\pi_i$ \KwonQN.

As
the name suggests, the string fluid behaves remarkably
like a continuum of noninteracting Nambu-Goto strings \GibbonsHF\SenKD.
This prompted the speculation that the string fluid
represents seed for formation of closed string as a
coherent state of unstable open strings \GibbonsHF\StromingerPC\GutperleXF.
Irrespective of such
effort, on the other hand, a more conservative interpretation
is possible in the classical limit. Note that the total energy
of an unstable D-brane scales as $\sim 1/g_s$, so with finite
electric field, the amount of
fundamental string charge per unit volume is enormous.
This fact allows a macroscopic interpretation of the
classical string fluid as a collection of large number
of long winding strings \GibbonsHF\SenXS. At the end of decay process,
nothing but closed strings would be left in full string
theory, so this is interpretation is both natural and quite
satisfying.

Once this macroscopic interpretation of string fluid is
accepted, it still remains to attribute the energy and
momentum of tachyon
matter. Again in real string theory, the only degrees of
freedom at the end are closed strings, so it is tempting
to view tachyon matter again as large number of highly
massive closed string modes. From inspection of low
energy dynamics, furthermore, it is suggested that these
high frequency modes of strings are not independent of
winding strings but rather must be on top of
the latter. Here, one must imagine the closed strings
carrying both the winding mode and additional
high frequency oscillation modes with the latter moving up
and down along the length of strings \SenXS.

Note that the conserved electric flux $\pi_i$ is different from
the electric field strength $F_{0i}$ in general.
Assuming a static configuration, $P_i=0 $, and also
$\pi_i\partial_i T=0$ for the sake of simplicity, we have
the relationship \GibbonsTV\KwonQN,
\eqn\sd{\pi_i={\cal H}F_{0i}}
and
\eqn\td{\pi_T={\cal H}\sqrt{1-\sum(F_{0i})^2},}
which also tells us that $\sum(F_{0i})^2$ lies
between 0 and 1. Interpreting the conserved charge $\pi_i$
as the string charge, one finds that this fluid description would be
consistent with the closed string picture if we
have the following ratio,
\eqn\e{{{\hbox{string charge}
\over \hbox{string energy}}}=e,}
for individual closed strings \SenXS, where  a new notation is
introduced for the size
of electric field strength
\eqn\e{e\equiv\sqrt{\sum(F_{0i})^2}.}
This picture has been tested by comparing the
fluid motion (of low energy open string theory) to
motion of many oscillatory Nambu-Goto strings \YeeEC.
Both static and dynamics properties have been seen to
match precisely.

We arrive at the picture of final state of unstable D-brane
decay as a large collection of aligned and stretched fundamental
strings with some high frequency modes turned on. Motion of string fluid,
encoded in $\pi_i$, represents low frequency undulation of this collection,
while the tachyon matter captures the energy-momentum of high frequency
modes moving up and down the same strings. In other words, the fluid picture
in the low energy open string theory is just a coarse-grained view
of  some specific type of states involving
many closed strings. This remarkable emergence of
closed string picture may provide some hint on relationship between
open strings and closed strings, and has been referred to as
an open/closed duality \SenXS.
The purpose of this note is to elevate this open/closed dual
picture of the final state of D-brane decay to a fully stringy level.

For this, we will
ask  how does the boundary state look like in the late time.
One way to answer this is to analyze the decay product of the
process, since we should be able to infer some properties of the
boundary state by looking at irradiation of closed strings.
In particular, contributions
to the decay arises dominantly from the disk one point function,
one can directly read off the composition of the late time boundary
state from the distribution of emitted closed string modes.
It is argued that in general one must consider multi-point disk
amplitude to see complete decay of the brane, except the case of
unstable D0. However, as we will see below, if one considers
unstable D1-brane wrapped on a circle, it turns out that one
point amplitude suffices in a manner quite similar to the case of
unstable D0 case in noncompact spacetime \LambertZR.

\newsec{Boundary state}
Our conventions are as follows. We set $\alpha'=1$ and define
the mode expansion as
\eqn\one{\eqalign{X^i_L(z)&= x_L^i -{i\over 2}p_L^i \ln(z) + {i\over
\sqrt{2}} \sum_{n\neq 0} {1\over n} a_n^i z^{-n},\cr
X^i_R(\bar z)&= x_R^i -{i\over 2}p_R^i \ln(\bar z) + {i\over
\sqrt{2}} \sum_{n\neq 0} {1\over n} \bar a_n^i \bar z^{-n} .}}
Where the worldsheet is a cylinder parameterized by $z=e^{\tau-i\sigma}$.
The commutation relations for spacelike bosons are given by
\eqn\two{\eqalign{[x_L^i, p_L^j]&=i \delta^{ij}, \quad [a_n^i,a_m^j]= n
\delta^{ij} \delta_{m+n},\cr
[x_R^i, p_R^j]&=i \delta^{ij}, \quad [\bar a_n^i,\bar a_m^j]= n
\delta^{ij} \delta_{m+n}.}}
If a bosonic coordinate $X^i$ is compactified on a circle of radius $R_i$, the
left and right momenta take the form
\eqn\three{k^i_L={n^i\over R_i}+ w_i R_i, \quad k^i_R= {n^i\over R_i}- w_i R_i,}
where $n^i\in Z$ is the KK momentum and $w_i\in Z$ is the winding
number in the $i$-th direction.

A boundary state imposes the boundary conditions for open strings
(ending on a D-brane) on the closed string modes. The boundary state
is useful since the sourcing of the closed string fields by the
D-brane can be  read off easily. The boundary conditions along the
worldvolume directions of a D-brane in the presence of a magnetic
field $b_{ij}$ are given by (the boundary of the worldsheet is located
at $\tau=0$ on the cylinder).
\eqn\four{\Big(g_{ij} \partial_n X^j + i b_{ij} \partial_t X^j \Big)
\mid B\rangle=0,}
where $\partial_n = \partial_\tau, \partial_t =\partial_\sigma$.
Inserting the mode expansion \one\ then gives the following condition
on the oscillators and zero modes
\eqn\five{\eqalign{\Big(  (g+b)_{ij} a_n^j+ (g-b)_{ij} \bar a_{-n}^j \Big) \mid
B\rangle&=0, \quad n\in Z\cr
\Big(  g_{ij} (p^j_L+p_R^j)+ b_{ij} (p_l^j-p_R^j)\Big) \mid
B\rangle&=0}.}
In the following we will focus on two (Euclidean) directions $i=1,2$
compactified on circles of radius $R_1,R_2$ respectively. The metric
metric and magnetic field given by
\eqn\six{g_{ij} =\pmatrix{1&0\cr0&1}, \quad b_{ij}= \pmatrix{0&-e\cr
e&0} .}
In this case the boundary state takes the form
\eqn\seven{\mid B \rangle = C \sqrt{1+e^2} \exp\big( -\sum_{n>0}
{1\over n} a_{-n}^i
 M_{ij} \bar a_{-n}^{j}\big) \mid B\rangle_0 .}
Where $C$ is a $e$ independent normalization factor and
\eqn\eight{M_{ij} = {(g-b)\over (g+b)} _{ij} = {1\over 1+e^2}
\pmatrix { 1-e^2 & 2e \cr -2e & 1-e^2}}
is an orthogonal matrix.

The zero mode part $\mid B\rangle_0$ of the boundary state which
solves \five\ is given by
\eqn\nine{\eqalign{\mid B\rangle_0&= \sum_{w_1,w_2} \mid k_L^1 = eR_2 w_2+ R_1
w_1, k_R^1= e R_2 w_2 - R_1w_1,\cr
&\quad\quad\quad\quad   k^2_L= -e R_1 w_1+ R_2w_2, k^2_R=-e R_1
w_1-R_2w_2 \rangle.}}
In the limit $R_1\to \infty$ only the zero mode sector $w_1=0$
survives. Note however that the winding $w_2$ along the $X^2$
 modifies the momentum in
the $X^1$ direction. For notational simplicity we will drop the index
on  the
winding number $w_2$ and the circle radius $R_2$ and replace them by
$w$ and $R$ respectively.

\medskip
A open string field $X(\sigma)$ is located at $\tau=0$. The presence
of the magnetic field modifies the OPE
\eqn\ten{X^i(\sigma_1)X^j(\sigma_2) = -{2\over 1+e^2} \delta^{ij} \ln
| e^{i\sigma_1} -e^{i\sigma_2}| + 2 {e\over 1+e^2} \epsilon^{ij}
  \epsilon(\sigma_1,\sigma_2). }
In particular a boundary vertex operator of the form $e^{i \sqrt{1+e^2} X^1}$
has conformal dimension one. Indeed this operator can be used to
define an exactly
marginal deformation of the boundary state \seven.
\eqn\twelv{\mid B(\lambda)\rangle = \exp\Big( -2\pi \lambda \oint
{d\sigma\over 2\pi} e^{i \sqrt{1+e^2} X^1(\sigma)} \Big) \mid B
\rangle.}
The part of the boundary state which does not involve any oscillators
in the $X^1$ direction can be evaluated by simply expanding the
boundary deformation in $\lambda$ and using \ten\ to evaluate the
correlators,
\eqn\thirte{\mid B(\lambda)\rangle_0= \sum_n {(-2\pi \lambda )^n\over
n!} \prod_{i=1}^n \oint {d\sigma_ i \over 2\pi} \prod_{i<j} |
e^{i\sigma_i}-e^{i\sigma_j}|^2 e^{i n \sqrt{1+e^2}  x_1}   \mid
B\rangle.}
Using the fact
\eqn\fourteen{\prod_{i=1}^n \oint {d\sigma_ i \over 2\pi} \prod_{i<j} |
e^{i\sigma_i}-e^{i\sigma_j}|^2 = n!.}
One can express \thirte\ as
\eqn\fifte{\mid B(\lambda)\rangle_0 = {1\over 1+ \lambda e^{i
\sqrt{1+e^2}  x_1} } \mid B\rangle .}

An analytic continuation $X^1\to i X^0$ turns the spacelike coordinate
$X^1$ into a timelike coordinate $X^0$. The boundary state stays real
if one also continues $e\to -i e$, this analytic continuation
corresponds to turning on an electric field in the $0,2$ directions
instead of a magnetic field in the $1,2$ direction.
After analytic continuation to the timelike case \fifte\
becomes
\eqn\sixte{\mid B(\lambda)\rangle_0 = {1\over 1+ \lambda e^{
\sqrt{1-e^2}  x_0} } \mid B\rangle .}
An alternative
calculation of the boundary state of a rolling tachyon in a background
electric field, involving a T-duality on $R$, a boost and another
T-duality was given in \ReyXS\ and gives the same result.

The origin of the matrix  $M$ in \seven\ can be understood more
clearly in the latter description: It merely reflects the Lorentz
transformation on the pair $(X^0_L+X^0_R, X_L^2-X_R^2)$ since the
T-dualization maps the electric field $e$ to the velocity $e$.
Therefore, it is clear the oscillator part of the boundary state
looks more or less the same as  the case with $e=0$, once we use
the co-moving coordinates in the T-dual description. In particular
this implies that the oscillator part of the boundary state is as
simple as before, if we concentrate on the part involving
oscillators from 25 spatial directions in the co-moving frame of
T-dual description. The latter can then be expressed in the
following way:
\eqn\eightt{\mid B(\lambda)\rangle_0 =  {i \over 2} \sum_{w}  \int
d{\cal E} {(2\pi \lambda)^{i{\cal E}} \over \sinh \pi {\cal E}}
\mid p^0_L =p^0_R = \sqrt{1-e^2} {\cal E} + eR w, p^2_L=-p^2_R= R
w\rangle ,}
where we used
\eqn\sevet{{1\over 1+ \lambda e^{
 x_0} } = {i \over 2} \int d{\cal E} {(2\pi
\lambda)^{i{\cal E}} \over \sinh \pi {\cal E}} e^{i {\cal E} x_0} . }


\newsec{Decay of D1 with electric flux}

The decay of the unstable brane into closed string modes has been
discussed in \LambertZR. Here, let us concentrate on the case of D1.
The basic object is the one point function on
the disk of
a closed string state $\langle {\cal E}, w, N\mid$, with energy
${\cal E}$  winding number $w$ and left-right symmetric transverse
oscillator excitations of level $N$,
\eqn\ninet{\eqalign{U({\cal E}_s,w, k, N) &= \langle {\cal E}_s, w,
N\mid B\rangle
= \sqrt{1-e^2} {(2\pi
\lambda)^{i{\cal E}}}{1 \over \sinh \pi {\cal E}}  }.}
Where ${\cal E}$ appearing on the right hand side  of \ninet\ is related to the
closed string energy ${\cal E}_s$ by
\eqn\twent{{\cal E} ={1\over \sqrt{1-e^2}}\big( {\cal E}_s - e R w\big) }
and ${\cal E}_s$ satisfies the usual mass shell condition
\eqn\twone{-{\cal E}_s^2 + (w R)^2+k_\perp^2 +4(N-1)=0.}
The total number of closed string states and the total energy produced
in the decay of the
brane is then given by
\eqn\twtwo{{\bar N\over V} = \sum_{states} {1\over 2{\cal E}_s} |
U({\cal E}_s)|^2,
\quad {\bar E\over V}= \sum_{states} | U({\cal E}_s)|^2,}
where the sum goes over all on shell closed string states.
Recall that, for a given level $N$, the number of closed string
states
with left and right oscillators matched
grows exponentially in critical bosonic string theory for large level $N$ as
 \eqn\twthree{c(N)= {1\over \sqrt{2}}N^{-{27/4}} e^{4\pi \sqrt{N} }.}
In the original calculation of the one point function in \LambertZR\ a
gauge was used \HwangAQ\EvansQU\ that  sets all timelike oscillators of
the closed string states to zero. One important question is whether the
presence of the electric field changes the large $N$ estimate \twthree\
of the sum over states \twtwo.

As far as the closed strings are concerned the boundary electric field
is equivalent to a constant antisymmetric tensor background. For the
non-zero mode excitations this leads to a chiral orthogonal rotation
of right movers with respect to the left movers, resulting from the
Lorentz transformation on $(X^0_L+X^0_R,X^2_L-X^2_R)$. The obvious
choice of the gauge is then to adopt the same timelike gauge as
in \LambertZR , except that now the timelike oscillators in the T-dual
co-moving coordinate system are set to zero. This not only kills
complications due to timelike oscillators in Sen' boundary state,
as in \LambertZR , but also removes possible extra $e$-dependence
that could have arisen from the matrix $M$. In the co-moving frame,
$e$-dependence of $M$ is absorbed by the oscillator redefinition.\foot{
This gauge choice is also justified indirectly by the cylinder
amplitude of the next section.}

This way, the only net effect of
$e$ in summing over the states will come from the zero mode part
multiplied by simple exponential factor  \twthree. With this,
the total radiated energy for large $N$ behaves as
\eqn\twfour{{\bar E\over V} \sim R(1-e^2)\sum_{w}  \int dk_\perp^{24}
\sum_N N^{-27/4}
e^{ 4\pi \sqrt{N} } e^{ - {2 \pi\over \sqrt{1-e^2}} \big(  \sqrt{
(wR)^2+k_\perp^2 + 4(N-1)}- eR w\big)},}
where we restored the normalization factor of $R$.
In the sector with no winding $w=0$, \twfour\ is exponentially
suppressed because of the electric field. This implies that the amount
of energy radiated away into unwound closed string is negligible
\MukhopadhyayEN\NagamiYZ.

However if the winding modes are taken into account
the picture changes drastically. Again with large $N$, the
winding sectors such that
\eqn\twfive{wR\simeq 2 {e
\sqrt{N} \over \sqrt{1-e^2}} + \xi,}
with small $\xi$ can be seen to contribute without exponential
suppression. Expanding the exponent in \twfour
\eqn\twsix{\eqalign{&4\pi \sqrt{N}  - {2 \pi\over \sqrt{1-e^2}} \big(  \sqrt{
(wR)^2+k_\perp^2 + 4(N-1)}- eR w\big)\cr \cr = &  -{\pi \over 2
\sqrt{N}} \big(
(1-e^2)\;\xi^2+ k_\perp^2\big)+o({1\over N}) }.}
It is easy to see that the Hagedorn growth is cancelled by the
exponential suppression coming from $e^{-2\pi {\cal E}}$.
In the spirit of large $N$ limit, we will replace the sums over
$w$ and $N$ to integrals
\eqn\si{R\sum_{w} \sum_N
\quad\to\quad   \int d(wR) \int dN,}
which in turn gives,
\eqn\sii{\sum_{w} \int dk_\perp^{24} \sum_N=
\int {1\over 2}\,{\cal E}_s d {\cal E}_s \int dk_\perp^{24}
\int d(wR) .}
Recall that along the saddle points
\eqn\mass{{\cal E}_s \simeq \sqrt{(wR)^2+4(N-1)}\simeq {2\sqrt{N}\over
\sqrt{1-e^2}},}
up to higher order corrections in $1/N$.
Then, the integrals become
\eqn\finali{(1-e^2)\int {1\over 2}\,{\cal E}_s d{\cal E}_s \int d\xi
\int dk_\perp^{24}
N^{-27/4}e^{-\pi \big(
(1-e^2)\;\xi^2+ k_\perp^2\big) /2  \sqrt{N}}.}
Evaluating this, we find all factors of $N$
and $\sqrt{1-e^2}$ drops out, and a simple integral emerges
\eqn\final{{\bar E\over V}\, \sim \, \int d{\cal E}_s.}
Note that, with the winding mode
taken into account, the exponential suppression disappears and
the total energy becomes linearly divergent. This is the same
behavior as for the D0 brane in \LambertZR, and indicates that the
energy of the
decaying D1 branes (with electric field) goes mainly into highly
wound strings.

To make contact with the low energy result described in section 1,
let us note that the saddle point condition \twfive\hskip 1mm can be written in
terms of string energy as
\eqn\compare{wR\simeq  e{\cal E}_s,}
which gives
\eqn\eagain{{{\hbox{string charge}
\over \hbox{string energy}}}=e.}
To summarize, we found that decay of D1 proceed primarily by many
single string emissions and that individual strings emitted satisfy
the above condition. This is a strong evidence that the unstable D1
with electric fields become a collection of strings with winding
numbers in a manner precisely predicted by low energy approach.

As an aside, let us discuss one obvious consequence in the
decay process. Note that the width
of the Gaussian distribution for  $k_\perp$ is $N^{1/4}$,
so the typical string has transverse velocity,
\eqn\vel{{\sqrt{\langle k_\perp^2\rangle} \over {\cal E}_s}
\sim {N^{1/4} \over {\cal E}_s} \sim (1-e^2)^{1/4} {1\over \sqrt{{\cal E}_s}},}
which is suppressed by a factor of $(1-e^2)^{1/4}$ when
string charge dominates the energy. Furthermore, since the
winding energy is quantized in unit of $R$, the closed
strings emitted tend to be rather heavy
as ${\cal E}_s >|w||R|\ge|R|\gg 1$ when $R$ is large.
Although the unstable D-brane is disintegrated
into a number of such winding strings quickly, these
closed strings take long time in ``radiating" into the nearby
spacetimes. The suppression comes primarily from having
to radiate heavy strings with winding modes.

\newsec{Cylinder amplitudes}
By unitarity the total number of closed string states can be related
to the imaginary part of a cylinder amplitude constructed from the
boundary state. Although the open string interpretation and modular
transformation properties of the full
cylinder amplitude (including the real part)  are still unclear it is
useful to understand the absence of exponential suppression found in
section 3 from this perspective.
For the number of closed strings states emitted by a decaying brane one finds
\LambertZR,
\eqn\thtwo{\eqalign{{\bar N\over V}&=  \sum_{states} {1\over 2 {\cal E}_s} |
U({\cal E}_s)|^2
={\rm Im} \langle B \mid {b_0^+c_0^- \over 2(L_0+\bar L_0 -i
\epsilon)} \mid B\rangle}.}
Note that the dependence on the electric field $e$ in the cylinder
amplitude comes  purely from the zero mode contribution. This can be
understood from the fact that the chiral rotation, discussed in section
3, leaves $\bar L_0$  and hence the propagator on the right
hand side of \thtwo\ invariant.

The one point function is given by \ninet\
where ${\cal E}$ is given by \twent.
\eqn\thnine{\eqalign{{\bar N\over V}&= \sum_{states} \sum_{w} {1\over
2{\cal E}_s}
|U({\cal E}_s)|^2\cr
&= R(1-e^2) \sum_{states} \sum_{w} {1\over 2{\cal E}_s } {1\over  \pi
\sinh^2 \left( {{\cal E}_s+ e
R w\over \sqrt{1-e^2}}\right)}}.}
Where we separated the summation over the winding modes $w$ in the
$X^2$ direction,  from the
sum over all physical states. \thnine\ can be brought into a more
convenient form by using the  formulas
\eqn\ththree{ {1\over (\sinh \pi y )^2}= \sum_{n>0} 4 n e^{-2\pi
n y}}
and
\eqn\thfour{\eqalign{{1\over y} e^{-2\pi y n} &= {1\over \pi}
\int
dk_0 {1\over y^2 + k_0^2} e^{2\pi i k_0 n}\cr
&= {1\over \pi} \int dk_0 \int_0^\infty  dt e^{-t( k_0^2+y^2)}
e^{2\pi i k_0 n} .}}
Putting everything together  \thnine\ becomes
\eqn\thnine{\eqalign{{\bar N\over V}&=  {2 R\over \pi^2} \sqrt{1-e^2}
\sum_{states} \sum_w \sum_{n>0} n e^{ 2\pi n
{e R w\over \sqrt{1-e^2}}} \int dk_0 \int dt e^{- t(k_0^2+ {{\cal E}_s^2
\over 1-e^2})} e^{2\pi i k_0 n}}.}
Where ${\cal E}_s$ is given by \twone. The Gaussian integrals over $k_0$ and
the twenty four transverse momenta $k_\perp$ give
\eqn\fourty{{\bar N\over V} = {2R\over \pi}(1-e^2) \sum_{states} \sum_w
\sum_{n>0}  n e^{ 2\pi n
{e R w\over \sqrt{1-e^2}}} \int d\tilde t \tilde t ^{-25/2} e^{- {\pi
n^2 \over  \tilde t (1-e^2)}} e^{- \tilde t ( (wR)^2 + 4(N-1))} ,}
where we have rescaled  $t = \pi (1-e^2) \tilde t $.
The sum
over winding modes  can be rewritten using the Poisson resummation formula
\eqn\fttwo{\sum_k e^{-\pi a k^2 + 2\pi i b k}= {1\over a^{1\over 2}} \sum_m
e^{-{\pi (m-b)^2\over a}} .}
The summation over transverse string oscillators  produces
\eqn\ftthree{{\bar N\over V} = {2\over \pi  } (1-e^2)\sum_{k,n>0} n
\int {d\tilde
t\over \tilde t^{13}} { e^{{4 \pi  \tilde t}}\over  \prod_m (1-
e^{-{2 \pi \tilde t m}})^{24}} e^{- {\pi \over \tilde t}
{n^2\over 1-e^2}} e^{- {\pi^2\over \tilde t} {1\over R^2} ( k - { i e
R n\over \sqrt{1-e^2}} )^2}. }
A modular transformation $s={1\over  2 \tilde t}$ expresses \ftthree\
as a open string partition function.
\eqn\ftfour{{\bar N\over V} = {2^{13}\over \pi }  (1-e^2)  \sum_{n>0,k} \int
{ds\over s} {1\over \prod_m (1-e^{-\pi s m})^{24}} e^{-\pi s \big( -1+
n^2 + {k^2\over R^2}\big)}.}
Note that the absence of an exponential suppression in the
ultraviolet ($s\to \infty$) region of the integral since the
exponentially growing term in the exponent of \ftfour\ is cancelled
by the $n=1$ term precisely. Note that the
cylinder amplitude also gives an indication that analysis given in
section 3 is correct. A electric field dependent change in  \twtwo\
and \twthree,  that changes the saddle point calculation given in
section 3,  should also invalidate
the cancellation in \ftfour, and this does not happen.

\newsec{Summary}

We studied closed string one point amplitude on the boundary state of
unstable D-branes. Wrapping a D1 on a circle, and turning on electric
field along the circle, we find that closed strings produced by the
decay are primarily given by certain winding strings with a definite
ratio between the winding energy and the oscillator energy and that
all energy of the initial system is accounted for by this decay channel.

This indicates that the final state of unstable D-brane in the
tree-level approximation is really made of closed strings with winding
number, in accordance with previous expectation based on low energy
approaches. In terms of languages of the latter, the combined
system of string fluid and tachyon matter is really a macroscopic
collection of stretched fundamental strings with high frequency
oscillations moving along the length.
While the one-point amplitude is most telling for case of unstable D1,
we expect that the same closed string interpretation will work for
higher dimensional D-branes.
In the large radius limit, this also implies that produced
strings are extremely heavy and do not easily disperse into the neighboring
spacetime.

\vskip 1cm
\centerline{\bf Acknowledgment }
Both authors are grateful to the theory group at CERN for
hospitality during the `Strings at CERN' workshop.
 The work of MG is supported in part by NSF grant PHY-0245096.
 Any opinions,
findings and conclusions expressed in this material are those of the
authors and do not necessarily reflect the views of the National
Science Foundation.

\listrefs

\end